\documentclass[aip,reprint,aps]{revtex4-1}
\usepackage{graphicx}
\usepackage{dcolumn}
\usepackage{bm}
\usepackage{color}
\usepackage[colorlinks=true,linkcolor=green,anchorcolor=green,citecolor=green,urlcolor=green]{hyperref}
\usepackage{amsthm,amsmath,amssymb}

\draft 

\begin{document}


\title{Interlayer magnetic interactions in $\pi/3$-twisted bilayer CrI$_3$} 
\author{Haodong Yu}
\affiliation{ School of Physical Science and Technology $\&$ Key Laboratory for Magnetism and Magnetic Materials of the MoE, Lanzhou University, Lanzhou 730000, China.}
\affiliation{ Lanzhou Center for Theoretical Physics and Key Laboratory of Theoretical Physics of Gansu Province, Lanzhou University, Lanzhou 730000, China}

\author{Jize Zhao}
\email{zhaojz@lzu.edu.cn}
\affiliation{ School of Physical Science and Technology $\&$ Key Laboratory for Magnetism and Magnetic Materials of the MoE, Lanzhou University, Lanzhou 730000, China.}
\affiliation{ Lanzhou Center for Theoretical Physics and Key Laboratory of Theoretical Physics of Gansu Province, Lanzhou University, Lanzhou 730000, China}

\author{Fawei Zheng}
\email{fwzheng@bit.edu.cn}
\affiliation{ Centre for Quantum Physics, Key Laboratory of Advanced Optoelectronic Quantum Architecture and Measurement (MOE), School of Physics, Beijing Institute of Technology, Beijing, 100081, China.}
\affiliation{ Beijing Key Lab of Nanophotonics $\&$ Ultrafine Optoelectronic Systems, School of Physics, Beijing Institute of Technology, Beijing, 100081, China.}

\date{\today}

\begin{abstract}
The interlayer magnetic interaction in bilayer CrI$_3$ plays a crucial role for its device applications. In this work, we studied the interlayer magnetic interaction in $\pi/3$-twisted bilayer CrI$_3$ using first-principles calculations.  Our calculations show that the interlayer coupling can be ferromagnetic or antiferromagnetic depending crucially on lateral shift. The strongest antiferromagnetic interlayer interaction appears in the $\bar{A}A$-stacking. The magnetic force theory calculations demonstrate that such an antiferromagnetic interaction is dominanted by the $e_g$-$e_g$ channel. Particularly, the interlayer antiferromagnetic interaction is very sensitive to external pressure. This highly tunable interlayer interaction makes $\pi/3$-twisted bilayer CrI$_3$ a potential building block for magnetic field effect transistors and pressure sensors.
\end{abstract}


\maketitle 


CrI$_3$ monolayer is one of the firstly discovered atomically thin two-dimensional magnetic materials.\cite{HuangXu2017} Due to its high magnetic anisotropy,\cite{ChoiKim2020} CrI$_3$ monolayer remains magnetically ordered up to 45 K.\cite{HuangXu2017} Its properties can be tuned by strain,\cite{Yangyan2019,WuYuan2019,VishkayiA2020} charge,\cite{Baiya2018,Z2018} and external field,\cite{GuoWu2018,Behera2019,Ghosh2019} which makes it an ideal platform for the study of spintronics. Recent experiments have been focusing on CrI$_3$ bilayers, which are promising candidates for device applications. The interlayer coupling in CrI$_3$ bilayer peeled from CrI$_3$ bulk is antiferromagnetic~(AFM) while the intralayer coupling remains ferromagnetic~(FM).\cite{JiangM2018,HuangX2018,SeylerX2018,KleninJ2018,GudelliG2019,KimT2019,SunW2019,GuoP2020} The interlayer AFM coupling can be flipped to FM by external magnetic field, and the flipping brings a huge transition of electric resistence.\cite{WangM2018,SongX2018} Such bilayer structure with huge magnetoresistance is a natural field effect transistor~(FET). Therefore, the interlayer AFM coupling in CrI$_3$ bilayer is an essential prerequisite to such device applications. Detailed theoretical studies found that the interlayer magnetic interactions depend on the stacking order,\cite{SivadasX2018,JiangJ2019,JangH2019,XiaoT2021}
such as bilayer CrI$_3$ with lateral shift,\cite{SunW2019,SivadasX2018}, bilayer CrI$_3$ with twist,\cite{XiaoT2021,AkramB2021,Gibertini2021}, and heterogenous bilayer CrI$_3$/CrGeTe$_3$.\cite{ShangK2020}  Due to the symmetry of the lattice, the largest twist angle is $\pi/3$. One may expect that some interesting phenomena can emerge under such twist and thus it calls for an immediate study.

\begin{figure}[tbp]
\centering
\includegraphics[width=8.7 cm]{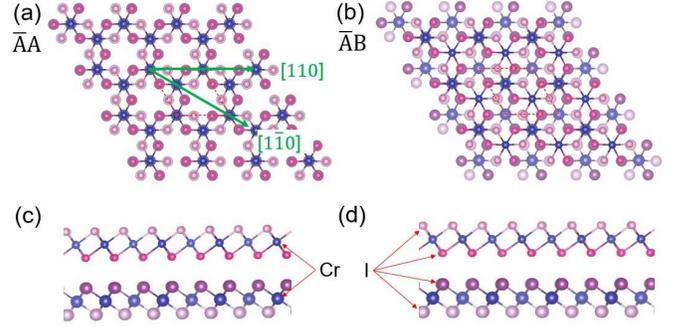}
\caption{\label{fig1}
The top (a) and side (c) views of $\bar{A}A$-stacked CrI$_3$ bilayer, and that (b,d) of $\bar{A}B$-stacked bilayer. The red dash parallelograms are unit cells, and the green vectors show the [100] and [1$\bar{1}$0] high-symmetry lines. The Cr and I atoms are shown in different colors and sizes for different positions.
}
\end{figure}

The Cr atoms in a CrI$_3$ monolayer form a honeycomb lattice, which is composed of two sublattices. The honeycomb lattice is similar to the structures of graphene\cite{NoboselovF2004} and $h$-BN monolayer.\cite{NagashimaO1995} However, these three materials are different in symmetry. The point symmetry group of graphene is $D_{6h}$. The $\pi/3$-rotation, reflection and inversion are all its symmetry operations. Each operation switches the two equivalent sublattices. Considering the high-symmetry of graphene, the bilayer graphene has only a few stacking orders, such as $AA$-stacking and $AB$-stacking. On the other hand, the symmetry of $h$-BN monolayer ($D_{3h}$ group) is lower than that of graphene. It gives more different bilayer stacking orders. The symmetry of CrI$_3$ monolayer is also lower than that of graphene, but it is higher than that of  $h$-BN monolayer. The corresponding group is $D_{3d}$. The symmetry operations include reflection and inversion, which switch the two equivalent sublattices. The $\pi/3$-rotation can also switch the two sublattices. However, considering the positions of I atoms, the system after $\pi/3$-rotation is different from its original structure. It means that the $\pi/3$-rotation is no longer a symmetry operation. Therefore, the $AA$ and $AB$ stackings in bilayer graphene correspond to four different stacking orders in bilayer CrI$_3$, which are $AA$, $\bar{A}A$, $AB$, and $\bar{A}B$ stackings. The $\bar{A}A$ and $\bar{A}B$ stackings are shown in Fig.~\ref{fig1}.  And the difference between $\bar{A}A$  and $AA$ stackings are shown in supporting information, Fig.~S1. The interlayer magnetic interactions in $AA$- and $AB$-stacked bilayer CrI$_3$ have been extensively investigated in literature.\cite{SivadasX2018} However, the study of the interlayer interactions in a $\pi/3$-twisted bilayer CrI$_3$, such as $\bar{A}A$-stacking and $\bar{A}B$-stacking, are still missing. In the following, we will systematically study the interlayer magnetic interactoins in $\pi/3$-twisted bilayer CrI$_3$ as well as in $\pi/3$-twisted bilayer CrBr$_3$ and CrCl$_3$. We found that the AFM interlayer coupling in $\bar{A}A$-stacked bilayer CrI$_3$ is the strongest, while that in $\bar{A}A$-stacked bilayer CrBr$_3$ and CrCl$_3$ is relatively weaker. Moreover, these interlayer coupling is very sensitive to external pressures, suggesting their potential applications in FET and pressure sensor devices.

The density functional theory(DFT) calculations in this work were carried out by using the Vienna ab initio simulation package (VASP).\cite{KresseF1996} The exchange-correlation functional is subjected to the generalized gradient approximation (GGA) in the Perdew, Burke, and Ernzerhof (PBE) form.\cite{PerdewE1996} The van der Waals interaction correction was considered with the D3 method of Grimme et. al.\cite{Grimme1996}  A Hubbard correction with 3 eV of on-site Coulomb repulsion (U) was adopted in the Cr atoms. The atomic structures were fully relaxed until the residual forces were less than 0.1 meV/{\AA}, with the experiment lattice constants.\cite{McGuire2015} The reciprocal space integrations were performed on a $9\times 9\times 1$ Monkhorst-Pack grid. And the plane-wave cutoff energy  was 450 eV. A vacuum space larger than 15 {\AA} was used so that the interlayer coupling can be neglected.

The unit cell of CrI$_3$ monolayer contains two Cr atoms and six I atoms. Each Cr atom is at the center of an I-octahedron. The octahedral crystal field splits the high-energy $e_g$ and low-energy $t_{2g}$ orbitals. The three valence electrons of Cr$^{3+}$ fill in the $t_{2g}$ orbitals of one spin, showing an effective 3 $\mu B$ magnetic moment. And the Cr-I-Cr superexchange interaction favors the FM ground states.\cite{Z2018,WangW2016,KashinA2020} When we put another CrI$_3$ layer with $\pi/3$-rotation on top of a non-rotated CrI$_3$ layer, there are van der Waals interactions between the two layers. Each layer keeps in FM state, while the magnetic interactions between the two layers can be either FM or AFM type.

\begin{figure}[tbp]
\centering
\includegraphics[width=8.7 cm]{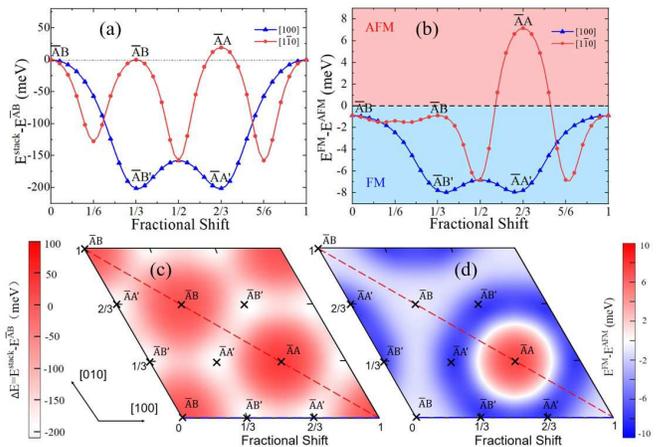}
\caption{\label{fig2}
The stacking energy (a,c) and interlayer exchange energy (b,d) of $\pi/3$-twisted bilayer CrI$_3$. Panel (a) and (b) are the calculation results for lateral shift along [100] and [1$\bar{1}$0] high symmetry lines, while panel (c) and (d) are the calculation results for the full two-dimensional lateral shifts.
}
\end{figure}

We firstly considered the lateral shifts of top layer CrI$_3$ along two high symmetry directions, which are [100] and [1$\bar{1}$0] directions. The initial structure is $\bar{A}B$-stacking. The $\bar{A}A$-stacking corresponds to a 2/3 fractional lateral shift along [1$\bar{1}$0] direction. The stacking energy, defined as the total energy variation of the stacking order, is used to show the stability of bilayers. The calculated stacking energies are shown in Fig.~\ref{fig2}(a), where the stacking energy of $\bar{A}B$-stacking is choosen to be zero. In order to get the relaxed bilayer structures with constrained lateral shifts, we fixed the in-plane positions of Cr atoms, and fully relaxed the off-plane coordinates of Cr atoms and all coordinates of I atoms. From the figure, we can see that the stacking energies of $\bar{A}A$ and $\bar{A}B$ stackings are at peak positions in the [1$\bar{1}$0] direction, and the $\bar{A}A$-stacking has the highest stacking energy. The local minimums of stacking energy appear at 1/6, 1/2, and 5/6 lateral shifts in the [1$\bar{1}$0] direction. The local minimums in [100] direction are at $\bar{A}B'$- and $\bar{A}A'$-stacking orders, which are at 1/3 and 2/3 shifts. The stacking energy of 1/2 shift in [100] direction is a local maximum, which has the same stacking energy of 1/2 shift in [1$\bar{1}$0] direction. Actually, they are equivalent stacking orders, and the 1/2 shift is a saddle point in two dimensional lateral shift. In order to see the landscape of stacking energies for full two dimensional lateral shift, we calculated the stacking energies on a $10\times 10$ grid of shifts. All the structures are fully relaxed with Cr-atom lateral shift constrained. The calculation results are shown in Fig.~\ref{fig2}(c). We can see that the high symmetry directions [100] and [1$\bar{1}$0] crossed all the local minimum and maximum points. The point which has the highest stacking energy in the whole two dimentional shift is $\bar{A}A$-stacking, while $\bar{A}B'$-stacking and $\bar{A}A'$-stacking are degenerated stacking orders with the lowest stacking energies.

\begin{figure}[tbp]
\centering
\includegraphics[width=8.5 cm]{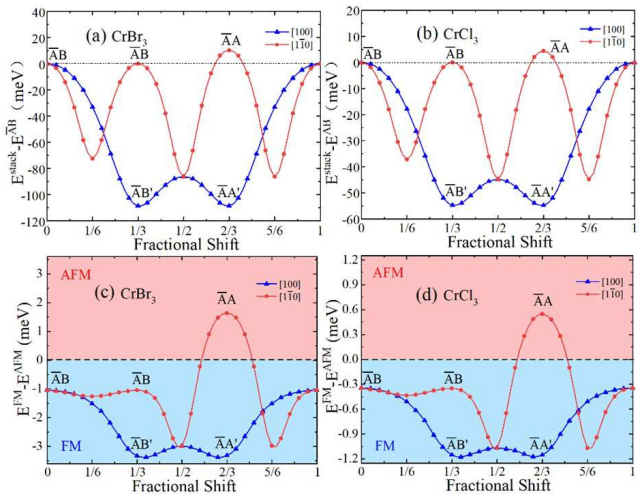}
\caption{\label{fig3}
The stacking energy (a,b) and interlayer exchange energy (c,d) of $\pi/3$-twisted CrBr$_3$ and CrCl$_3$ bilayers with lateral shift along [100] and [1$\bar{1}$0] high symmetry lines.
}
\end{figure}

The $\pi/3$-twisted bilayer CrI$_3$ may have ferromagnetic or antiferromagnetic interlayer interactions. The energy difference between the ferromagnetic and antiferromagnetic interlayer configurations is defined as the interlayer exchange energy. The calculated interlayer exchange energy with lateral shift along [100] and [1$\bar{1}$0] directions are shown in Fig.~\ref{fig2}(b). The negative interlayer exchange energy indicates the ferromagnetic configuration, while the positive interlayer exchange energy indicates the antiferromagnetic order. The plotted line in the direction [100] are all below zero, thus the bilayer CrI$_3$ with this kind of shift has ferromagnetic order. The $\bar{A}A'$-stacking and $\bar{A}B'$-stacking orders have the strongest ferromagnetic interlayer exchange energies, which are about -7.9 meV. Most of the plotted line in the direction [1$\bar{1}$0] is also below zero, while the data points around 2/3 shift have positive interlayer exchange energies. The point for $\bar{A}A$-stacking is at the peak position, and the associated value is as high as 7.1 meV, indicating the $\bar{A}A$-stacking has strong antiferromagnetic interlayer interaction. In order to see the landscape of interlayer exchange energies for all the two dimensional lateral shifts, we calculated them on the $10\times 10$ grid of shifts, which is shown in Fig.~\ref{fig2}(d). We can see that the antiferromagnetic interlayer interactions exist only around $\bar{A}A$-stacking, forming a disc. The $\bar{A}A$-stacking has the strongest antiferromagnetic interlayer interactions in all two dimensional lateral shifts. It is also one order larger than that of $AB'$-stacking, and about three times larger than that of $AB'_1$-stacking. Around the disc,  there are lateral shifts with strong ferromagnetic interlayer interactions, including $\bar{A}A'$- and $\bar{A}B'$-stacking orders.

The high stacking energy indicates that the $\bar{A}A$-stacking is energetically unfavorable. But we will show that it is still accessible. The interlayer interactions are weak van der Waals interactions. The calculated cleavage energy is about 27 meV per atom while the typical chemical bond energy is of the order of eV per atom. Thus, the interlayer interaction is much weaker than that of chemical bond. Therefore, the stacking structure can be tuned by external forces, such as the force from the tip of scanning tunneling miscroscope. Besides, the $\bar{A}A$-stacking may also exist in some local areas of twisted CrI$_3$ bilayer.

Besides CrI$_3$ bilayer, we also considered the $\pi/3$-twisted bilayer CrBr$_3$ and CrCl$_3$. The calculated stacking energies and interlayer exchange energies are shown in Fig.~\ref{fig3}. We can see that the main features are the same with that of CrI$_3$ bilayers. However, their absolute values are quite different. The stacking energies for $\bar{A}A$-stacking and $\bar{A}B'$-stacking of CrBr$_3$ bilayer are 10.3 meV and -108.5 meV, respectively. They are about half of that for CrI$_3$ bilayers. The stacking energies for $\bar{A}A$-stacking and $\bar{A}B'$-stacking of CrCl$_3$ bilayer are 4.4 meV and -54.7 meV, respectively. They are only about 1/4$\sim$1/5 of that for CrI$_3$ bilayers. The antiferromagnetic interlayer interaction of $\bar{A}A$-stacking is also much smaller in CrBr$_3$ and CrCl$_3$ bilayers, and they are 1.6 meV and 0.5 meV. They are about one order smaller than that of CrI$_3$ bilayer. The intelayer exchange energies for CrCl$_3$ bilayer are so small that we have to set the force criterion in relaxation as low as 0.1 meV/{\AA} to obtain the correct values. The stacking energy and interlayer exchange energy of CrBr$_3$ and CrCl$_3$ bilayers in full two dimensional lateral shifts are shown in supplementary materials.

\begin{table}
\centering
\caption{Hopping parameters between Wannier functions of Cr$_1$-I$_1$-I$_2$-Cr$_2$ chain in the fully relaxed $\bar{A}A$ stacked CrI$_3$ bilayer. The hopping parameters are in eV unit.}
\begin{tabular}{|c|c|c|c|c|}
\hline
\multicolumn{3}{|c|}{Cr$_1$-I$_1$,Cr$_2$-I$_2$} & \multicolumn{2}{c|}{I$_1$-I$_2$} \\
\hline
d$_{z^2}$-p$_z$ & \begin{tabular}[c]{@{}c@{}}d$_{yz}$-p$_y$\\d$_{xz}$-p$_x$\end{tabular} & others & \begin{tabular}[c]{@{}c@{}}p$_i$-p$_i$\\(i=x,y,z)\end{tabular} & \begin{tabular}[c]{@{}c@{}}p$_i$-p$_j$\\(i$\ne$ j;\\i,j=x,y,z)\end{tabular}  \\
\hline
1.04   & -0.43  & \begin{tabular}[c]{@{}c@{}}absolute value\\ $<$ 0.04\end{tabular} & -0.16$\sim$-0.18  & -0.10$\sim$-0.12 \\
\hline
\end{tabular}
\end{table}

\begin{figure}[tbp]
\centering
\includegraphics[width=8.5 cm]{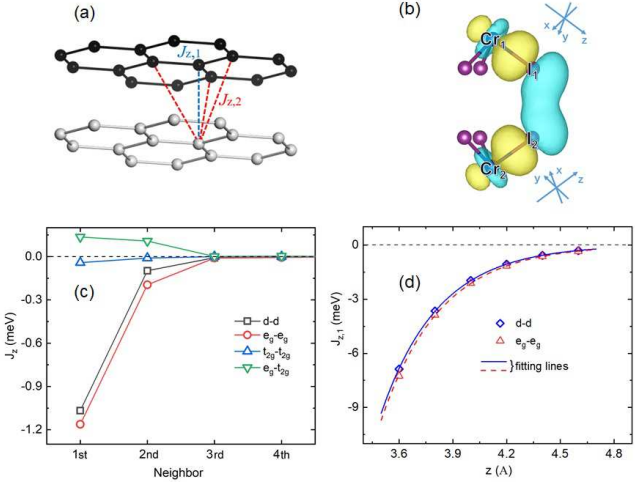}
\caption{\label{fig4}
(a) The schematic picture of interlayer magnetic interactions in $\bar{A}A$-stacked CrI3 bilayer. The first ($J_{z,1}$) and the second neighboring interlayer couplings ($J_{z,2}$) are represented by blue and red colors, respectively. The black and gray atoms represent Cr atoms in top and bottom CrI$_3$ layers, respectively.(b) The Cr$_1$-I$_1$-I$_2$-Cr$_2$ atom chain, and the isosurface of the superposition of $d_{z^2}$ and -$p_z$ orbitals. It shows the large overlap at $d_{z^2}$-$p_z$, and $p_z$-$p_z$, and explains the crucial role of $e_g$-$e_g$ channel. The Wannier functions for Cr$_1$ and I$_1$ are defined in the coordinate shown at top, while that for Cr$_2$ and I$_2$ are defined in coordinate shown at bottom. (c) The total and orbital decomposed $J_z$ values as a function of neighboring distance. The positive and negative values correspond to FM and AFM interactions, respectively. (d) The nearest interalayer exchange interactions as functions of interlayer distance.
}
\end{figure}

In order to find the origin of the strong antiferromagnetic interlayer interaction of $\bar{A}A$-stacked CrI$_3$ bilayer, we continue to construct the Hamiltonian of the system. The Heisenberg Hamiltonian for a bilayer CrI$_3$ system can be written as:
\begin{equation}
\begin{aligned}
H=&H_i+H_o\\
H_o=&  -\sum_{i,j}J_{i,j}\bm{S}_i^u\cdot\bm{S}_j^d \\
	=&-J_{z,1}\sum_{\langle{ij}\rangle}\bm{S}_i^u\cdot\bm{S}_j^d \\
	&-J_{z,2}\sum_{\langle\langle{ij}\rangle\rangle}\bm{S}_i^u\cdot\bm{S}_j^d +...
\end{aligned}
\end{equation}
The Hamiltonian is composed of intralayer interaction $H_i$ and the coupling between the two layers $H_o$. The $\bm{S}_i^u$ and $\bm{S}_j^d$ are the magnetic moments of the $i$-th and $j$-th Cr atoms for up and down CrI$_3$ layers, respectively. The magnetic moments on Cr atoms are all normalized to 1. And the $\langle{i,j}\rangle$ and $\langle\langle{i,j}\rangle\rangle$ are the nearest and the next-nearest interlayer neighbors of Cr atoms. The nearest and the next-nearest interlayer magnetic interactions in $\bar{A}A$-stacking CrI$_3$ bilayer are shown in Fig.~\ref{fig4}(a). The interlayer magnetic interactions were obtained by using the magnetic force theory (MFT) with the code TB2J,\cite{HeB2021} which is an open source Python package for the automatic computation of magnetic interactions between atoms. MFT has an unique and useful feature that it can decompose magnetic interactions into different orbitals. In our calculations, the interlayer magnetic interactions are decomposed into $e_{g}$-$e_{g}$, $t_{2g}$-$t_{2g}$ and $e_{g}$-$t_{2g}$ channels. The orbitals are defined in local coordinate systems, which are different for top and bottom CrI$_3$ layers as shown in Fig.~\ref{fig4}(b). The calculated results are summarized in Fig.~\ref{fig4}(c). We can see that the magnetic interactions decrease quickly with increasing distances. The interaction of the third nearest neighbor is already very weak. Both $e_{g}$-$e_{g}$ and $t_{2g}$-$t_{2g}$ interactions are antiferromagnetic, while the $e_{g}$-$t_{2g}$ interaction is ferromagnetic. The absolute values of $t_{2g}$-$t_{2g}$ and  $e_{g}$-$t_{2g}$ interactions are much smaller than that of $e_{g}$-$e_{g}$. Thus, the total $d$-$d$ magnetic interaction is mainly contributed by $e_{g}$-$e_{g}$ interaction.

The interlayer magnetic interactions of CrI$_3$ bilayer are super-super exchange type. The magnetic interaction between two Cr atoms in the two layers can be conducted via several Cr-I-I-Cr paths. We then calculated the maximally localized Wannier functions of the $\bar{A}A$-stacked CrI$_3$ bilayer to find the detailed magnetic interaction paths. The Wannier functions for Cr$_1$, I$_1$, I$_2$, and Cr$_2$ atoms are calculated in their local coordinates as shown in Fig.~\ref{fig4}(b). In our calculations, the maximally localized wannier functions and the Hamiltonian parameters were obtained by using Wannier90 code.\cite{MostofiM2014} The hopping parameters between these orbitals are shown in Table I.  The hopping parameters for $d_{z^2}$-$p_z$, $d_{yz}$-$p_y$, and $d_{xz}$-$p_x$ are large, which are 1.04 eV, -0.43 eV, and -0.43 eV, respectively. However, the absolute values for the other $d$-$p$ hopping parameters are less than 0.04 eV due to their different symmetries. And all the $p$-$p$ hopping parameters between I$_1$ and I$_2$ are of the same order. The parameters for $p_x$-$p_x$, $p_y$-$p_y$, and $p_z$-$p_z$ are in the range from -0.18 eV to -0.16 eV, while the parameters for the other $p$-$p$ hoppings are in the range from -0.12 eV to -0.10 eV, which are one third weaker. The Cr-$e_g$ orbitals consisit of $d_{z^2}$ and $d_{x^2-y^2}$ orbitals. The $d_{x^2-y^2}$ orbital only has very small hopping parameters with the I-$p$ orbitals. Therefore, the magnetic interaction in Cr$_1$-I$_1$-I$_2$-Cr$_2$ path is mainly contributed by the orbital path of $d_{z^2}$-$p_z$-$p_z$-$d_{z^2}$. The isosurface plot of the superposition of $d_{z^2}$ and -$p_z$ orbitals are shown in Fig.~\ref{fig4}(b), which clearly shows the large overlap between these orbitals. The weakest part of the orbital path is the interlayer $p_z$-$p_z$ hopping. It can be tuned by interlayer distances. The radial functions of atomic orbitals are in the form of exponential functions times polynomials, then the $p_z$-$p_z$ hopping parameters are exponentially decreases with increasing the interlayer distances. Therefore, the $e_g$-$e_g$ magnetic interaction is also supposed to decrease exponentially with increasing interlayer distances. The calculated nearest-neighbor magnetic interaction $J_{z,1}$ as a function of interlayer distance is shown in Fig.~\ref{fig4}(d). The data can be perfectly fitted by formula $J_{z,1}=-e^{a(b-z)}$ meV, where $z$ is the interlayer distance, and $a,b$ are two fitting parameters. The fitted results are $a=3.10$ \AA$^{-1}$ and $b=4.22$ \AA. The total d-d magnetic interactions can also be fitted by formula $J_{z,1}=-e^{a(b-z)}$ meV. The fitted results for $d$-$d$ interactions are $a=3.03$ \AA$^{-1}$ and $b=4.25$ \AA. We can see that the decrease of interlayer distances, which can be applied by external pressure, increases the antiferromagnetic interactions dramatically. Thanks to this peculiar feature, a FET based on $\bar{A}A$-stacked CrI$_3$ bilayer can also be used as a pressure sensor.  Supposing that the FET is in the conductive state, the magnetic field has already flipped all the magnetic moments of CrI$_3$ bilayer into the same direction. Then, an external pressure can switch off the electric current by changing the system back to antiferromagnetic.

In conclusion, we studied the interlayer interactions of $\pi/3$-twisted bilayer CrI$_3$. For each two-dimensional lateral shift, the stacking energy and interlayer exchange energy are calculated after atom structure relaxation. Our calculations show that most of the lateral shifts have ferromagnetic interlayer interactions, while the antiferromagnetic interlayer interaction appears only in  $\bar{A}A$-stacking and the nearby shifts. The interlayer interactions in $\pi/3$-twisted bilayer CrBr$_3$ and CrCl$_3$ also show the same properties, except that their stacking and interlayer magnetic interactions are relatively weak. We performed the MFT calculations to study the strong antiferromagnetic interaction in $\bar{A}A$-stacked CrI$_3$, and found that it is mainly contributed by $e_{g}$-$e_{g}$ channel via $d_{z^2}$-$p_z$-$p_z$-$d_{z^2}$ path. It is exponentially dependent on the interlayer distance, and is very sensitive to external pressure. Based on these results, a pressure sensor is designed.

\section{Acknowlegements}
This work was supported by National Natural Science Foundation of China (Grants Nos. 12022415, 11974056, 11874188, 12047501). The data that supports the findings of this study are available within the article and its supplementary material.


\begin{thebibliography}{99}

\bibitem{HuangXu2017}
    B. Huang, G. Clark, E. Navarro-Moratalla, D. R. Klein, R. Cheng, K. L. Seyler, D. Zhong, E. Schmidgall, M. A. McGuire, D. H. Cobden, W. Yao, D. Xiao, P. Jarillo-Herrero, and X. Xu,
  \href{https://doi.org/10.1038/nature22391}{Nature} {\bf 546}, 270 (2017).

\bibitem{ChoiKim2020}
    Y. Choi, P. Ryan, D. Haskel, J. McChesney, G. Fabbris, M. A. McGuire and J.-W. Kim,
  \href{https://doi.org/10.1063/5.0012748}{Appl. Phys. Lett.} {\bf 117}, 022411 (2020).

\bibitem{Yangyan2019}
    B. Yang, X. Zhang, H. Yang, X. Han, and Y. Yan,
  \href{https://doi.org/10.1063/1.5091958}{Appl. Phys. Lett.} {\bf 114}, 192405 (2019).

\bibitem{WuYuan2019}
    Z. Wu, J. Yu, and S. Yuan,
  \href{https://doi.org/10.1039/C8CP07067A}{Phys. Chem. Chem. Phys.} {\bf 21}, 7750 (2019).

\bibitem{VishkayiA2020}
    S. I. Vishkayi, Z. Torbatian, A. Qaiumzadeh, and R. Asgari,
  \href{https://doi.org/10.1103/PhysRevMaterials.4.094004}{Phys. Rev. Materials} {\bf 4}, 094004 (2020).

\bibitem{Baiya2018}
    S. Baidya, J. Yu, and C. H. Kim,
  \href{https://doi.org/10.1103/PhysRevB.98.155148}{Phys. Rev. B}  {\bf 98}, 155148 (2018).

\bibitem{Z2018}
    F. Zheng, J. Zhao, Z. Liu, M. Li, M. Zhou, S. Zhang, and P. Zhang,
  \href{https://doi.org/10.1039/C8NR03230K}{Nanoscale} {\bf 10}, 14298 (2018).

\bibitem{GuoWu2018}
    G. Guo, G. Bi, C. Cai, and H. Wu,
  \href{https://doi.org/10.1088/1361-648X/aac96e}{J. Phys.: Condens. Matter} {\bf 30}, 285303 (2018).

\bibitem{Behera2019}
    A. K. Behera, S. Chowdhury, and S. R. Das,
  \href{https://doi.org/10.1063/1.5096782}{Appl. Phys. Lett.} {\bf 114}, 232402 (2019).

\bibitem{Ghosh2019}
    S. Ghosh, N. Stoji\v{c}, and N. Binggeli,
  \href{https://doi.org/10.1016/j.physb.2019.06.040}{Phys. B} {\bf 570}, 166 (2019).

\bibitem{JiangM2018}
    S. Jiang, J. Shan, and K. F. Mak,
  \href{https://doi.org/10.1038/s41563-018-0040-6}{Nat. Mater.} {\bf 17}, 406 (2018).

\bibitem{HuangX2018}
    B. Huang, G. Clark, D. R. Klein, D. MacNeill, E. Navarro-Moratalla, K. L. Seyler, N. Wilson, M. A. McGuire, D. H. Cobden, D. Xiao, W. Yao, P. Jarillo-Herrero, and X. Xu,
  \href{https://doi.org/10.1038/s41565-018-0121-3}{Nat. Nanotechnol.} {\bf 13}, 544 (2018).

\bibitem{SeylerX2018}
    K. L. Seyler, D. Zhong, D. R. Klein, S. Gao, X. Zhang, B. Huang, E. Navarro-Moratalla, L. Yang, D. H. Cobden, M. A. McGuire. D. H. Cobden, D. Xiao, W. Yao, P. Jarillo-Herrero, and X. Xu,
  \href{https://doi.org/10.1038/s41567-017-0006-7}{Nat. Phys.} {\bf 14}, 277 (2018).

\bibitem{KleninJ2018}
    D. R. Klein, D. Macneill, J. L.  Lado, D. Soriano, E. Navarro-Moratalla, K.  Watanabe, T. Taniguchi , S. Manni, P. Canfield, J. Fern\'{a}ndez-Rossier, and P. Jarillo-Herrero,
  \href{https://doi.org/10.1126/science.aar3617}{Science} {\bf 360}, 1218 (2018).

\bibitem{GudelliG2019}
    V.K. Gudelli, and G. Guo,
  \href{https://doi.org/10.1088/1367-2630/ab1ae9}{New J. Phys.} {\bf 21}, 053012 (2019).

\bibitem{KimT2019}
    H. H. Kim, B. Yang, S. Li, S. Jiang, C. Jin, Z. Tao, G. Nichols, F. Sfigakis, S. Zhong, C. Li, S. Tian, D. G. Cory, G. X. Miao, J. Shan, K. F. Mak, H. Lei, K. Sun, L. Zhao, and A. W. Tsen,
  \href{https://doi.org/10.1073/pnas.1902100116}{Proc. Natl. Acad. Sci. USA} {\bf 116}, 11131 (2019).

\bibitem{SunW2019}
    Z. Sun, Y. Yi, T. Song, G. Clark, B. Huang, Y. Shan, S. Wu, D. Huang, C. Gao, Z. Chen, M. McGuire, T. Cao, D. Xiao, W.-T. Liu, W. Yao, X. Xu, and S. Wu,
  \href{https://doi.org/10.1038/s41586-019-1445-3}{Nature} {\bf 572}, 497 (2019).

\bibitem{GuoP2020}
    K. Guo, B. Deng, Z. Liu, C. Gao, Z. Shi, L. Bi, L. Zhang, H. Lu, P. Zhou, L. Zhang, Y. Cheng, and B. Peng,
  \href{https://doi.org/10.1007/s40843-019-1214-y}{Sci. China Mater.} {\bf 63}, 413 (2020).

\bibitem{WangM2018}
     Z. Wang, I. Guti\'{e}rrez-Lezama, N. Ubrig, M. Kroner, M. Gibertini, T. Taniguchi, K. Watanabe, A. Imamo\v{g}lu, E. Giannini, and A. F. Morpurgo,
  \href{https://doi.org/10.1038/s41467-018-04953-8}{Nat. Commun.} {\bf 9}, 2516 (2018).

\bibitem{SongX2018}
     T. Song, X. Cai, M. W.-Y. Tu, X. Zhang, B. Huang, N. P. Wilson, K. L. Seyler, L. Zhu, T. Taniguchi, K. Watanabe, M. A. McGuire, D. H. Cobden, D. Xiao, W. Yao, and X. Xu,
  \href{https://doi.org/10.1126/science.aar4851}{Science} {\bf 360}, 1214 (2018).

\bibitem{SivadasX2018}
     N. Sivadas, S. Okamoto, X. Xu, C. J. Fennie, and D. Xiao,
  \href{https://doi.org/10.1021/acs.nanolett.8b03321}{Nano Lett.} {\bf 18}, 7658 (2018).

\bibitem{JiangJ2019}
     P. Jiang, C.Wang, D. Chen, Z. Zhong, Z. Yuan, Z.Y. Lu, and W. Ji,
  \href{https://doi.org/10.1103/PhysRevB.99.144401}{Phys. Rev. B} {\bf 99}, 144401 (2019).

\bibitem{JangH2019}
     S. W. Jang, M. Y. Jeong, H. Yoon, S. Ryee, and M. J. Han,
  \href{https://doi.org/10.1103/PhysRevMaterials.3.031001}{Phys. Rev. Materials} {\bf 3}, 031001(R) (2019).

\bibitem{XiaoT2021}
     F. Xiao, K. Chen, and Q. Tong,
  \href{https://doi.org/10.1103/PhysRevResearch.3.013027}{Phys. Rev. Res.} {\bf 3}, 013027 (2021).

\bibitem{AkramB2021}
     M. Akram, H. LaBollita, D. Dey, J. Kapeghian, O. Erten, and A. S. Botana,
  \href{https://doi.org/10.1021/acs.nanolett.1c02096}{Nano Lett.} {\bf 21}, 6633 (2021).
  
\bibitem{Gibertini2021}
    M. Gibertini, \href{https://doi.org/10.1088/1361-6463/abc2f4}
    {J. Phys. D: Appl. Phys.} {\bf 54}, 064002 (2021).
  
\bibitem{ShangK2020}
     J. Shang, X. Tang, X. Tan, A. Du, T. Liao, S. C. Smith, Y. Gu, C. Li, and L. Kou,
  \href{https://doi.org/10.1021/acsanm.9b02055}{ACS Appl. Nano Mater.} {\bf 3}, 1282 (2020).

\bibitem{NoboselovF2004}
     K. S.Novoselov, A. K. Geim, S. V. Morozov, D. Jiang, Y. Zhang,  S. V. Dubonos, , I. V. Grigorieva, and  A. A. Firsov,
  \href{https://doi.org/10.1126/science.1102896}{Science} {\bf 306}, 666 (2004).

\bibitem{NagashimaO1995}
    A. Nagashima, N. Tejima, Y. Gamou, T. Kawai, and C. Oshirna,
  \href{https://doi.org/10.1103/PhysRevLett.75.3918}{Phys. Rev. Lett.} {\bf 75}, 3918 (1995).

\bibitem{KresseF1996}
    G. Kresse, and J. Furthm{\"u}ller,
  \href{https://doi.org/10.1103/PhysRevB.54.11169}{Phys. Rev. B} {\bf 54}, 11169 (1996).

\bibitem{PerdewE1996}
    J. P. Perdew, K. Burke, and M. Ernzerho,
  \href{https://doi.org/10.1103/PhysRevLett.77.3865}{Phys. Rev. Lett.} {\bf 77}, 3865 (1996).

\bibitem{Grimme1996}
    S. Grimme, J. Antony, S. Ehrlich, and H. Krieg,
  \href{https://doi.org/10.1063/1.3382344}{J. Chem. Phys.} {\bf 132}, 154104 (2010).

\bibitem{McGuire2015} M. A. McGuire, H. Dixit, V. R. Cooper, and B. C. Sales,
  \href{https://doi.org/10.1021/cm504242t}{Chem. Mater.} {\bf 27}, 612 (2015)

\bibitem{WangW2016}
    H. Wang, F. Fan, S. Zhu, and H. Wu,
  \href{https://doi.org/10.1209/0295-5075/114/47001}{Europhys. Lett.} {\bf 114}, 47001 (2016).

\bibitem{KashinA2020}
    I. V. Kashin, V. V. Mazurenko, M. I. Katsnelson, and A. N. Rudenko,
  \href{https://doi.org/10.1088/2053-1583/ab72d8}{2D Mater.} {\bf 7}, 025036 (2020).

\bibitem{HeB2021}
    X. He, N. Helbig, M. J. Verstraete, and E. Bousquet,
  \href{https://doi.org/10.1016/j.cpc.2021.107938}{Comput. Phys. Commun.} {\bf 264}, 107938 (2021).

\bibitem{MostofiM2014}
    A. A. Mostofi, J. R. Yates, G. Pizzi, Y.-S. Lee, I. Souza, D. Vanderbilt and N. Marzari,
  \href{https://doi.org/10.1016/j.cpc.2014.05.003}{Comput. Phys. Commun.} {\bf 185}, 2309 (2014).


\end{thebibliography}
\end{document}